\documentclass[ conference, letterpaper, 10pt, times]{IEEEtran}
\IEEEoverridecommandlockouts
\usepackage{cite}
\usepackage[T5]{fontenc}
\usepackage[utf8]{inputenc}
\usepackage{amsmath,amssymb,amsfonts}
\usepackage{graphicx}
\usepackage{textcomp}
\usepackage{xcolor}
\usepackage{graphicx}
\usepackage{tabularx}
\usepackage{multirow}
\DeclareRobustCommand*{\IEEEauthorrefmark}[1]{%
  \raisebox{0pt}[0pt][0pt]{\textsuperscript{\footnotesize #1}}%
  }
\usepackage{algorithm2e} % Load the algorithm2e package with ruled option
\RestyleAlgo{ruled} % You can also restyle it here
\usepackage{array} % To improve the table spacing if needed
\def\BibTeX{{\rm B\kern-.05em{\sc i\kern-.025em b}\kern-.08em
    T\kern-.1667em\lower.7ex\hbox{E}\kern-.125emX}}
\begin{document}

\title{PSO-based Sliding Mode Current Control of Grid-Forming Inverter in Rotating Frame\\

\thanks{\textit{Corresponding authors: V.H. Bui (vhbui@umich.edu) and W. Su (wencong@umich.edu)}}
}

\author{
    \IEEEauthorblockN{Quang-Manh Hoang\IEEEauthorrefmark{1},
    {Guilherme Vieira Hollweg}\IEEEauthorrefmark{1}, 
    Akhtar Hussain\IEEEauthorrefmark{2},
    Sina Zarrabian\IEEEauthorrefmark{3},
    Wencong Su\IEEEauthorrefmark{1,}\IEEEauthorrefmark{*}, 
    Van-Hai Bui\IEEEauthorrefmark{1,}\IEEEauthorrefmark{*}}
    
    \IEEEauthorblockA{\IEEEauthorrefmark{1}Department of Electrical and Computer Engineering, University of Michigan-Dearborn, USA}
    
    \IEEEauthorblockA{\IEEEauthorrefmark{2}Department of Electrical and Computer Engineering, Laval University, Canada}
    
    \IEEEauthorblockA{\IEEEauthorrefmark{3}Department of Electrical Engineering, State University of New York (SUNY), Maritime College, USA}
    
}

\maketitle

\begin{abstract}
The Grid-Forming Inverter (GFMI) is an emerging topic that is attracting significant attention from both academic and industrial communities, particularly in the area of control design. The Decoupled Average Model-based Sliding Mode Current Controller (DAM-SMC) has been used to address the need such as fast response, fixed switching frequency, and no overshoot to avoid exceeding current limits. Typically, the control parameters for DAM-SMC are chosen based on expert knowledge and certain assumptions. However, these parameters may not achieve optimized performance due to system dynamics and uncertainties. To address this, this paper proposes a Particle Swarm Optimization (PSO)-based DAM-SMC controller, which inherits the control laws from DAM-SMC but optimizes the control parameters offline using PSO. The main goal is to reduce chattering and achieve smaller tracking errors. The proposed method is compared with other metaheuristic optimization algorithms, such as Genetic Algorithm (GA) and Simulated Annealing (SA). Simulations are performed in MATLAB/Simulink across various scenarios to evaluate the effectiveness of the proposed controller. The proposed approach achieves a substantial reduction in convergence time, decreasing it by 86.36\% compared to the GA and by 88.89\% compared to SA. Furthermore, the tracking error is reduced by 11.61\% compared to the conventional DAM-SMC algorithm. The robustness of the proposed method is validated under critical conditions, where plant and control model parameters varied by up to 40\%.
\end{abstract}


\section{Introduction}
The integration of distributed energy resources (DERs) into power systems is increasing, primarily through inverter-based resources (IBRs), as part of the energy transition. Among these, the Grid-Forming Inverter (GFMI) has emerged as a critical technology that functions as a voltage source, capable of independently generating stable voltage and frequency, even in weak grid conditions. The modeling and control design of GFMI play essential roles in ensuring system stability and resilience, particularly within the voltage and current control loops \cite{khan2022grid, 9637882, 8586162,li2024stability}.

Some research in the literature has used a single-loop voltage controller to increase the bandwidth of voltage control. This is because, in a cascade structure, the outer loop of a classical controller may need to be slower than the inner loop \cite{9380975}. However, this approach may not ensure safe operation, as the current can exceed its limit, potentially causing the inverter to operate in grid-following mode during faults. Other studies have focused on the current control loop, using methods such as predictive current control \cite{luo2017predictive} and conventional PI control \cite{hoang2024decoupled}. However, the conventional PI controller may cause overshoot, which makes current limitation challenging, and may also result in poor transient and steady-state responses. Predictive current control, as implemented in \cite{luo2017predictive}, requires an accurate model, which is difficult to obtain when GFMI parameters are unavailable due to security concerns or imprecision caused by aging.

Another nonlinear control approach applied in both single-phase and three-phase inverters is sliding mode control (SMC) \cite{hollweg2022feasibility, 8668431, hoang2024decoupled, 4012138, 720325, CHAKRABARTY2016143}. Using SMC can increase system robustness with certain structures \cite{CHAKRABARTY2016143}, however, the chattering phenomenon is one of the main challenges of the SMC scheme. This issue arises due to factors such as the control order, structural complexity, sampling and switching frequency, and variations in plant model parameters \cite{levant2003higher}. Additionally, its control law is typically designed based on the converter switching model and hysteresis control type \cite{4012138}, which also contributes to high chattering and non-fixed switching frequency problems \cite{720325}. To address these drawbacks, high-order SMC has been proposed to mitigate the chattering phenomenon \cite{ullah2021power}; however, this introduces greater complexity in controller mathematical modeling and increases computational costs. On the other hand, DAM-SMC with a simple control structure is proposed in \cite{hoang2024decoupled}, which guarantees a constant switching frequency and consequently reduces the chattering phenomenon. However, in that study, the SMC parameters were often chosen based on expert knowledge with assumptions related to the control parameters between the direct and quadrature axes. Furthermore, the proposed method in \cite{hoang2024decoupled} does not account for changes in system parameters, which can lead to control performance that does not meet expectations.

To overcome the aforementioned drawbacks, a PSO-based DAM-SMC is proposed in this study. The DAM-SMC control laws are derived from \cite{hoang2024decoupled}, which has fixed switching frequency, and low computational cost. The control parameters are optimized offline using an effective PSO algorithm, with the cost function being the tracking error represented by the integral absolute error (IAE). The control parameter boundaries are determined based on expert knowledge, and the optimized control parameters found at the end of the optimization process are used to simulate the system in MATLAB/Simulink. The proposed PSO-based DAM-SMC is compared with the method proposed in \cite{hoang2024decoupled}, as well as with optimized GA-based and SA-based DAM-SMC. In this study, each algorithm is run 10 times, and the mean value of the IAE is obtained to ensure the reliability of the three optimization algorithms. The proposed method is simulated across various scenarios to validate its dynamic response, robustness, and convergence time. The proposed PSO-based DAM-SMC demonstrates superior performance. It features a simple SMC structure, fixed switching frequency, and low computational cost. Additionally, it significantly reduces the chattering phenomenon and tracking error by approximately 11.67\% compared to \cite{hoang2024decoupled}. Furthermore, the convergence time of PSO is reduced by 86.36\% compared to GA and 88.89\% compared to SA.

The reminder of the paper is organized as follows. The system configuration, including both the plant model and control structure, is presented in Section II. Following that, the proposed PSO-based DAM-SMC is introduced in Section III, along with the procedure for obtaining the optimized control parameters. In Section IV, the proposed method is validated through simulations in MATLAB/Simulink under various scenarios and compared with the DAM-SMC control scheme, GA-based DAM-SMC, and SA-based DAM-SMC. Moreover, upcoming works are also presented in this section. Finally, conclusions are presented in Section V.
\section{System configuration}
\begin{figure}[t]
\centerline{\includegraphics[scale = 0.6]{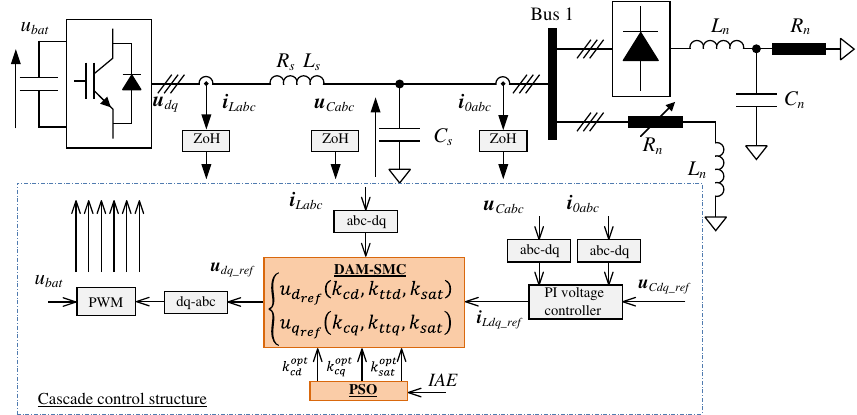}}
\caption{Configuration of the studied system.}
\label{fig_config}
\end{figure}

In this section, the configuration of the studied system as outlined in Figure. \ref{fig_config}. It consists of an inverter-based resource operating as a grid-forming inverter, along with both linear and nonlinear loads. In addition to supplying power to the linear load, the GFMI also energizes the nonlinear load. This serves as a test scenario for evaluating current control performance in the presence of significant harmonic currents.

Figure. \ref{fig_config} also shows the control structure featuring the cascade control scheme. In this section, we briefly outline the procedure for modeling the studied system and designing the DAM-SMC control, based on \cite{hoang2024decoupled}.
\subsection{Modeling studied system}
As mentioned previously, the studied system consists of a three-phase LC-filter inverter operating as a GFMI, an RL linear load, and a three-phase full-bridge diode rectifier connected to an RLC nonlinear load.

In the plant model, the switching model is used to mathematically implement the operation of three-phase inverter, as shown in (\ref{eq_Switch}). Where $\mathbf{SS}=\begin{bmatrix} SS_a & SS_b & SS_c \end{bmatrix}^T$ is the switching status of the inverter.
\begin{equation}
		\begin{bmatrix}\label{eq_Switch}
			u_{a}\\
			u_{b}\\
			u_{c}\\
		\end{bmatrix}= \dfrac{1}{3}
		\begin{bmatrix}
			2 & -1 & -1\\
			-1 & 2 & -1\\
			-1 & -1 & 2\\
		\end{bmatrix}
            \begin{bmatrix}
			SS_a\\
			SS_b\\
			SS_c\\
		\end{bmatrix}
            u_{bat}
	\end{equation}

To facilitate controller design, the Park transformation is used to convert the signals from the abc axis to the rotating dq coordinate system as in (\ref{eq2}) and (\ref{eq3}).
    \begin{equation}	\begin{bmatrix}\label{eq2}
			x_{\alpha}\\
			x_{\beta}\\
		\end{bmatrix}
		=\dfrac{2}{3}
		\begin{bmatrix}
			1 & \dfrac{-1}{2} & \dfrac{-1}{2}\\
			0 & \dfrac{\sqrt{3}}{2} & \dfrac{-\sqrt{3}}{2}\\
		\end{bmatrix}\begin{bmatrix}
			x_{a}\\
			x_{b}\\
			x_{c}
		\end{bmatrix}
	\end{equation}
	
	\begin{equation}
		\begin{bmatrix}\label{eq3}
			x_{d}\\
			x_{q}
		\end{bmatrix}=\begin{bmatrix}
			\cos{\theta} & \sin{\theta}\\
			-\sin{\theta} & \cos{\theta}
		\end{bmatrix}\begin{bmatrix}
			x_{\alpha}\\
			x_{\beta}\\
		\end{bmatrix}
	\end{equation}
 where $x$ generally denotes the considered variables with the corresponding voltage components and $\theta=\arctan{(\dfrac{u_{\beta}}{u_{\alpha}})}$.
 
 When transforming from the abc coordinate to the dq coordinate, coupling components appear in both the voltage and current components, as presented in (\ref{eq_Induc}) and (\ref{eq_Capa}), where $w$ is the angular frequency of the three-phase electrical system. 
        \begin{equation}
		\begin{bmatrix}\label{eq_Induc}
			i_{Ld}\\
			i_{Lq}
		\end{bmatrix}=\dfrac{1}{L_ss+R_s}
		\begin{bmatrix}
			u_d-u_{Cd}+i_{Lq}\omega{}L_s\\
			u_q-u_{Cq}-i_{Ld}\omega{}L_s
		\end{bmatrix}
	\end{equation} 
        \begin{equation}
		\begin{bmatrix}\label{eq_Capa}
			u_{Cd}\\
			u_{Cq}
		\end{bmatrix}=\dfrac{1}{C_ss}
		\begin{bmatrix}
			i_{Ld}-i_{0d}+u_{Cq}\omega{}C_s\\
			i_{Lq}-i_{0q}-u_{Cd}\omega{}C_s
		\end{bmatrix}
	\end{equation}
On the other hand, the RL linear load and nonlinear load are connected to the GFMI via bus 1, with the mathematical model detailed in \cite{hoang2024decoupled}.

\subsection{Cascade control design}
\subsubsection{Conventional PI voltage controller}
In this paper, the cascade control structure is implemented for inner loop control of GFMI where the voltage control loop is designed based on conventional PI controller on the dq frame. By compensating the load currents $\mathbf{i}_{0dq}$ and the coupling terms $\mathbf{u}_{Cdq} \omega C_s$, the control laws can be generated using (\ref{eq_PI}). 
        \begin{equation}
        \resizebox{0.9\hsize}{!}{$
		\begin{bmatrix}\label{eq_PI}
			i_{Ldref}\\
			i_{Lqref}
		\end{bmatrix}=\left(
		K_{pv}+\dfrac{K_{iv}}{s}
		\right) \begin{bmatrix}
			u_{Cdref}-u_{Cd}\\
			u_{Cqref}-u_{Cq}
		\end{bmatrix}
        +\begin{bmatrix}
			i_{0d}-u_{Cq}\omega{}C_s\\
			i_{0q}+u_{Cd}\omega{}C_s
		\end{bmatrix}$}
	\end{equation} 
where the control parameters can be determined by pole placement as:
	\begin{equation}\label{eq_PI_Para}
		\begin{cases}
			K_{pv}=2\xi{}\omega_vC_s\\
			K_{iv}=\omega_v^2C_s
		\end{cases}
	\end{equation}
where $\xi$ is the damping factor and $\omega_v$ is the natural frequency of the voltage control loop. The choice of $\xi=1$ means the critical damping case and $\omega_v=\dfrac{2\pi}{T_{vres}}$ can force the capacitor voltages respond in the desired time $T_{vres}$. The actual signals $i_{Labc}, u_{Cabc}, i_{0abc}$ are sampled by Zero-Order-Hold (ZoH) with a specific sampling time before being transformed into dq signals in the rotating frame.

\subsubsection{Decoupled average model-based sliding mode current controller}
To effectively respond to rapidly and frequently changing current reference requirements, the current controller needs to have a short response time. To meet these challenges, a decoupled average model-based sliding mode current controller for the studied LC-filtered inverter is proposed in \cite{hoang2024decoupled}, with the control laws outlined in (\ref{eq_SMC}).
\begin{equation}
    \resizebox{1\hsize}{!}{$
    \begin{cases}
        u_{dref}= &-L_s\dfrac{k_{cd}}{k_{ttd}} \operatorname{sat} \left(\dfrac{S_d}{k_{sat}}\right)+L_s\dfrac{d(i_{Ldref})}{dt}  -\omega L_s i_{Lq} + R_s i_{Ld} + u_{Cd} \\
        u_{qref}= &-L_s\dfrac{k_{cq}}{k_{ttq}} \operatorname{sat} \left(\dfrac{S_q}{k_{sat}}\right)+L_s\dfrac{d(i_{Lqref})}{dt} + \omega L_s i_{Ld} + R_s i_{Lq} + u_{Cq}
    \end{cases}$
    }
    \label{eq_SMC}
\end{equation}

 However, the control parameters of that controller \((k_{sat}, k_{cd}, k_{cq})\) are determined using expert knowledge, and the control parameters in different frames are assumed to have the same values \((k_{cd}\) and \(k_{cq})\). All of these factors can decrease control performance, leading to an increase in the tracking error between the reference and actual signals. To improve the efficiency of the average model-based sliding mode controller, this paper proposes an optimized DAM-SMC for GFMI by using PSO. The main objective is to minimize the tracking error, represented by the IAE, which is discussed in more detail in the next section.

\section{Proposed PSO-based optimized siliding mode control}

\begin{algorithm}
    \caption{PSO-based optimization framework for DAM-SMC.}\label{alg:PSO}

    $t \gets  0$\;
    $i \gets  50$\;
    Randomly initialize $\boldsymbol{X(0)^i} = [\boldsymbol{k_{cd}^i(0)}; \boldsymbol{k_{cq}^i(0)}; \boldsymbol{k_{sat}^i(0)}]$\;
    
    \While {$t \leq 45$}{
        Calculate $\textbf{IAE}$ values (\ref{eq_IAE}) of all particles $\boldsymbol{IAE(t)^i}$\ by running Simulink model\;
        Update the personal best swarm positions that are the set of control parameters according to the smallest $\textbf{IAE}$ values of each particle $\textbf{i}$: $\boldsymbol{p_{best}(\textit{t})^i} = \boldsymbol{X^i}(\min(\boldsymbol{IAE(\textit{1:t})^i}))$\;
        Update the global best swarm positions that is the control parameters set according to the smallest value of IAE in entire $\textbf{IAE}$ maxtrix: $g_{best}(t) = \boldsymbol{X}(\min(\boldsymbol{IAE(\textit{1:t})}))$\;
        
        Update swarm velocity:\\
        $\boldsymbol{V^i(t+1)} = \boldsymbol{w_{in}}\boldsymbol{V^i(t)} + \boldsymbol{c_1r_1}(\boldsymbol{p_{\text{best}}}(t)^i - \boldsymbol{X^i(t)}) + \boldsymbol{c_2r_2}(g_{best}(t) - \boldsymbol{X^i(t)})$\;
        
        Update the control parameters as known as swarm positions:\\
        $\boldsymbol{X^i(t+1)} = \boldsymbol{X^i(t)} + \boldsymbol{V^i(t+1)}$\;
        
        $t \gets t + 1$\;
    }
    
\end{algorithm}

PSO is an effective metaheuristic optimization algorithm developed by Kennedy and Eberhart \cite{488968} for optimizing a wide range of functions by simulating the behavior of biological systems. In this proposed control scheme, three SMC control parameters \((k_{sat}, k_{cd}, k_{cq})\) need to be determined, and they are treated as particles. The optimization process is described in Algorithm. \ref{alg:PSO} with the aim of minimizing the cost function IAE, as shown in (\ref{eq_IAE}).

\begin{equation}\label{eq_IAE}
    \textit{IAE} = \int_{0}^{T} |e(t)| \, dt
\end{equation}
where \(e(t)\) is the tracking error of the control signal, in this case, \(e(t) = i_{Ldqref}(t) - i_{Ldq}(t)\). As can be seen in Algorithm.  \ref{alg:PSO}, the PSO first randomly generates an initial matrix swarm positions of particles \(\boldsymbol{X^i(0)}\) with a size of \(50 \times 3\). After that, the values of \(\textbf{IAE}\) are evaluated by running the MATLAB/Simulink model. In this step, \(\textbf{IAE}\) is a \(50 \times 1\) matrix. In the next step, the best solution from the beginning to the current iteration is updated and saved to the \(\boldsymbol{p_{\text{best}}(t)^i}\) matrix. Simultaneously, the smallest value of \(IAE\) in the \(50 \times t\) matrix is also updated and stored in \(g_{best}(t)\). In this study, the optimization process stops if the iteration number reaches 45. This specific maximum number of iterations is chosen so that the fitness value can be fairly compared across the same number of iterations. If the iteration number is less than 45, the term swarm velocity \(\boldsymbol{V^i(t+1)}\) and the values of the SMC controller parameters \((\boldsymbol{k_{\text{sat}}^i(t+1)}; \boldsymbol{k_{\text{cd}}^i(t+1)}; \boldsymbol{k_{\text{cq}}^i(t+1)})\) are calculated for the next iteration by following the rules in Algorithm. \ref{alg:PSO}. The process is then executed again until the iteration number reaches the maximum limit. At the end of optimization process, the optimized control parameters \(k_{sat}^{opt}, k_{cd}^{opt}, k_{cq}^{opt}\) are obtained and used to run the Simulink model to achieve the optimized results.

\section{Simulation results and discussions}
\subsection{Simulation scenarios}
        \begin{table}
		\caption{Parameters of the system}\label{tab_SysPara}
		\centering
		\begin{tabular}{p{1.5in} p{1in}}
			\hline\hline
			\textbf{Model parameters} \\ \hline
			Sampling period   & $T_{sam}=50\mu{}s$\\
			dc bus voltage  & $u_{bat}=700V$\\
			LC-filter  & $L_s=2.4mH$\\
			&  $C_s=15\mu{}F$\\
			&  $R_s=0.1\Omega$\\
			Switching frequency  & $f_s=10kHz$\\
			Nonlinear load & $L_n=1.8mH$\\
			& $C_n=2.2mF$\\
			& $R_n=460\Omega$\\
                & $u_f=0.8V$\\
			Reference voltage amplitude & $u_{amp}=300V$\\
			Reference angular frequency & $\omega{}=100\pi{}rad/s$\\ \hline
			\textbf{Control parameters}  & \\ \hline
			Current time response 	& $T_{cres}=0.5ms$\\
			Voltage time response 	& $T_{vres}=10ms$\\ 
               Upper limit $[k_{cd},k_{cq},k_{sat}]$             & $[1, 1, 0.001]$\\
            Lower limit $[k_{cd},k_{cq},k_{sat}]$            & $[2000, 2000, 15]$\\ \hline
                \textbf{PSO}  & \\ \hline
			Inertia range 	& $w_{in} = [0.1;1.1]$\\
                Self adjustment weight 	& $c_1 = 1.49$\\
                Group adjustment weight 	& $c_2 = 1.49$\\ \hline
                \textbf{GA}  & \\ \hline
			Number of generation 	& $t = 45$\\
                Population size         & $50$\\\hline
                \textbf{SA}  & \\ \hline
			Initial temperature 	& $100$\\
			Number of iterations 	& $t = 45$\\ 
             \hline
		\end{tabular}
	\end{table}
To examine the efficiency of the proposed controller, several simulation scenarios and other metaheuristic optimization algorithms are implemented in MATLAB/Simulink, with the system parameters defined in Table. \ref{tab_SysPara}. The hardware used is a 13th Gen Intel(R) Core(TM) i5-13500HX at 2.50 GHz, the software environment is MATLAB R2023a, and the operating system is Windows 11. To ensure the reliability of the three optimization algorithms in this study, each algorithm is run 10 times, and the mean value of IAE is obtained.

During the 0.7 s simulation time, all control schemes undergo three simulation scenarios. First, the GFMI is connected to the RL linear load, and there is a load change at 0.1 s. Next, the nonlinear load is connected while the linear load is disconnected at 0.2 s. Finally, the inductance and resistance is increased by 40\% at 0.5 s. The detailed discussions are provided in the next section.

\subsection{Performance comparison}

\begin{figure}[b]
\centerline{\includegraphics[scale=1]{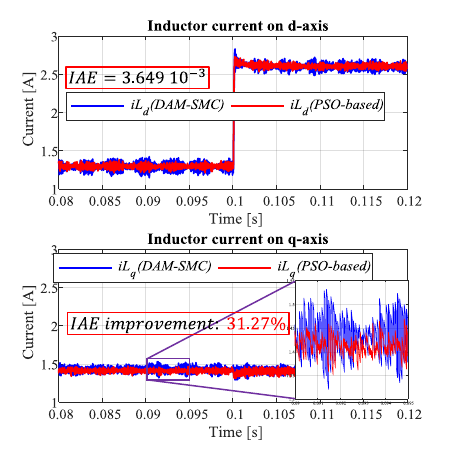}}
\caption{dq inductor currents under linear load change.}
\label{fig_iLdq_Linear}
\end{figure}

As mentioned above, the GFMI is first connected to the RL linear load, and a load change occurs at 0.1 s as shown in Figure. \ref{fig_iLdq_Linear}. In this situation, the superior control performance of the DAM-SMC is visually evident, with a fast response and approximately no overshoot. Furthermore, as shown in Figure. \ref{fig_iLdq_Linear}, the chattering phenomenon is significantly reduced when using the PSO-based DAM-SMC controller for the GFMI, compared to the \cite{hoang2024decoupled}. Additionally, the tracking error of the proposed method is approximately 31.27\% less than that of the normal DAM-SMC control scheme.

\begin{figure}[t]
\centerline{\includegraphics[scale=0.85]{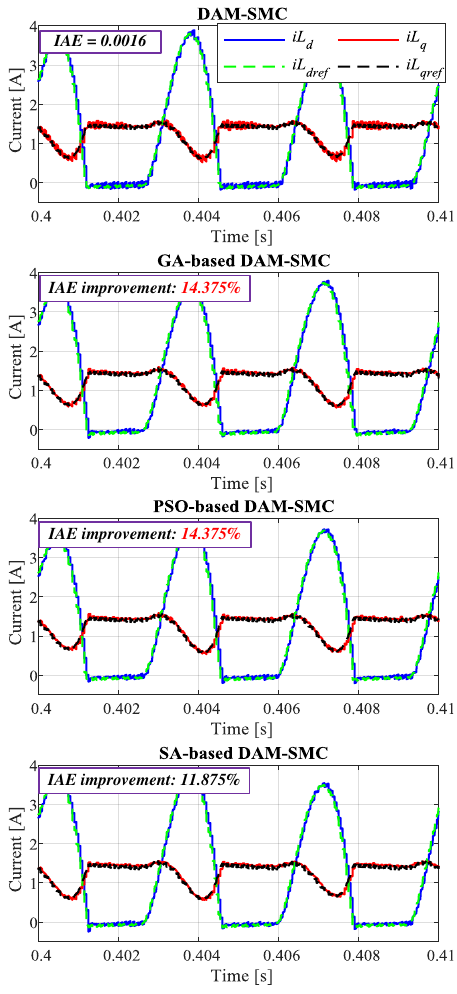}}
\caption{dq inductor currents under non-linear load.}
\label{fig_iLdq_NonLinear}
\end{figure}

\begin{figure}[t!]
\centerline{\includegraphics[scale=1]{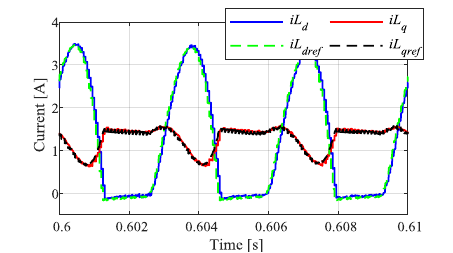}}
\caption{dq inductor currents under uncertainty.}
\label{fig_iLdq_Uncertain}
\end{figure}

In the second scenario, the nonlinear load is connected while the linear load is disconnected at 0.2 s. The dq inductor currents of four different control schemes between 0.4 s and 0.41 s are shown in Figure. \ref{fig_iLdq_NonLinear}. It is clear that the optimization-based approaches yield better results in terms of the chattering phenomenon, as represented by the IAE values of the four different control methods. Among these, the PSO-based DAM-SMC and GA-based DAM-SMC demonstrate good performance, with approximately 14.375\% improvement compared to the normal DAM-SMC. Finally, in the third scenario, a severe uncertainty is introduced into the model: variations in inductor parameters. The inductance and resistance are increased by 40\% at 0.5 s; this scenario did not appear in \cite{hoang2024decoupled}. The results shown in Figure. \ref{fig_iLdq_Uncertain} indicate the robustness of the PSO-based DAM-SMC. It is evident that the actual signals track the reference even in the presence of significant differences between the plant model parameters and the controller parameters.

\begin{figure}[t!]
\centerline{\includegraphics{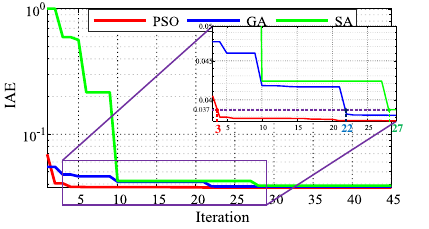}}
\caption{Fitness function for the optimization methods.}
\label{fig_Fitness}
\end{figure}

One interesting observation is that the control performance among the three optimization-based control schemes may look  the same without proper evaluation. However, Table. \ref{tab_IAE} shows a significant improvement in the PSO-based DAM-SMC controller in terms of IAE, with an 11.61\% improvement compared to the previous control scheme in \cite{hoang2024decoupled}. As shown in Figure. \ref{fig_Fitness} and Table. \ref{tab_IAE}, if we use IAE = 0.037 to compare convergence times, PSO reaches this value in just 3 iterations, approximately 86.36\% faster than GA (22 iterations) and 88.89\% faster than SA (27 iterations).
\begin{table}[t!]
    \caption{Performance comparison between four methods}
    \label{tab_IAE}
    \centering
    \begin{tabular}{m{0.7in} m{0.4in} m{0.4in} m{0.4in} m{0.4in}}
    \hline \hline
    Optimization Algorithms (OA) & Mean IAE & Improve-ment & Converge speed (iterations) \\ \hline
        GA & 0.08324 & 9.26\% & 22  \\
        PSO & \textcolor{red}{0.08108} & \textcolor{red}{11.61\%} & \textcolor{red}{3} \\
        SA & 0.0832 & 9.3\% & 27 \\
        Without OA \cite{hoang2024decoupled} & 0.09173 & & \\ \hline
    \end{tabular}
\end{table}

\section{Conclusion}
This study proposes a PSO-based DAM-SMC controller to improve the current control loop performance of GFMI. In which, all control parameters of DAM-SMC were optimized using the PSO algorithm. Simulation results demonstrate that the proposed PSO-based approach significantly outperforms others, including GA and SA, across various scenarios. Specifically, the tracking error was reduced by 11.61\% compared to the conventional DAM-SMC, and convergence time was shortened by 86.36\% compared to GA and 88.89\% compared to SA. As a future extension, we plan to implement the proposed control algorithm in a C-HIL simulation using the Typhoon HIL 604 and DSP TMS320F28335.
\section*{Acknowledgment}
The authors work was supported by the University of Michigan-Dearborn’s Office of Research “Research Initiation and Development".
\bibliographystyle{ieeetr}
\bibliography{Ref}

\end{document}